\batchmode
\documentclass[aps,prb,twocolumn,english]{revtex4-1}
\usepackage{amsmath,amssymb,bm}
\usepackage{graphicx}
\usepackage{color}
\usepackage{xspace}
\begin{document}

\title{Critical fields and fluctuations determined from specific heat and
magnetoresistance in the same nanogram SmFeAs(O,F) single crystal}

\author{S.Galeski{*}, P.W.J. Moll, N.Zhigadlo, K.Mattenberger, B. Batlogg}

\address{Solid State Physics Laboratory, ETH Zurich, CH-8093 Zurich, Switzerland }

\email{Corresponding author: galeskis@phys.ethz.ch}

\begin{abstract}
Through a direct comparison of specific heat and magneto-resistance
we critically asses the nature of superconducting fluctuations in
the same nano-gram crystal of SmFeAs(O, F). We show that although
the superconducting fluctuation contribution to conductivity scales
well within the 2D-LLL scheme its predictions contrast the inherently
3D nature of SmFeAs(O, F) in the vicinity {\normalsize{}$T_{c}$.
}Furthermore the transition seen in specific heat cannot be satisfactory
described either by the LLL or the XY scaling. Additionally we have
validated, through comparing H\textsubscript{c2} values obtained
from the entropy conservation construction (H\textsubscript{c2║ab}=-19.5
T/K and H\textsubscript{c2┴ab}=-2.9 T/K), the analysis of fluctuation
contribution to conductivity as a reasonable method for estimating
the $H{}_{c2}$ slope.
\end{abstract}
\maketitle

\section{Introduction}

The surprising discovery of superconductivity at 26K in LaFeAsO in
2008 was a beginning of a new era for superconductivity \citep{Kamihara2008}.
Soon after the initial discovery a great effort was taken to reach
higher transition temperatures, resulting in a discovery of dozens
of iron based superconductors with new compounds still being synthesized
\citep{Paglione2010}. However promising, the abundance of new structures
was not followed by the availability of high quality macroscopic samples.
In particular crystals of the '1111' family, with the highest $T_{c}=55~K$,
usually grow as flakes of $100-200~\mu m$ diameter, \citep{Zhigadlo2008}
making it challenging to study their bulk thermodynamics. In the face
of such difficulties newly discovered superconductors are traditionally
characterized by their transport properties. However the influence
of possible filamentary and surface superconductivity together with
the defect-induced vortex pinning the resistive transition tends to
be difficult to interpret. 

This is particularly visible when determining the $H{}_{c2}$ slope
from the typically smooth and featureless resistive data - depending
on the chosen criterion for the transition temperature (10, 50 90\%
of normal state resistivity) one might obtain$H_{c2}$ slopes differing
by more than a factor of 2-3 \citep{Lee2009}.

\textcolor{black}{One debate originating from these issues is the
discussion of dimensionality of the superconducting fluctuations in
SmFeAs(O, F) with reports of 2D and 3D behaviour \citep{Pallecchi,Welp2011}.
In the first work Palecchi et al. reported that the superconducting
fluctuation contribution to conductivity could be well parametrized
within the 2D-LLL scaling scheme in stark contrast to the second study
of Welp et al. who suggested the prevalence of 3D-LLL scaling of the
superconducting contribution to specific heat in fields up to 8T.
One possible explanation of this apparent discrepancy could be the
sample variability. On the other hand the analysis of scaling of the
superconducting fluctuation contribution to conductivity suffered
from the lack of high quality single crystals and was performed on
a polycrystalline sample making a proper analysis of fluctuation conductivity
very difficult. }

Here we measure \textit{on the same single crystal} of SmFeAs(O, F),
both heat capacity and resistivity near the superconducting transition
in fields up to 14T applied parallel and perpendicular to the FeAs
layers. Analysis of the phenomenology of the resistive transition
shows that the low temperature part of the transition is strongly
influenced by the vortex dynamics. On the other hand, the onset of
the transition can be well accounted for as originating from fluctuation
conductivity (with the same H\textsubscript{c2} slope as found in
specific heat ) and following the scaling form of the 2D Lowest Landau
Level (LLL) theory. The appearance of the specific heat anomaly accompanying
the transition also reveals a significant presence of fluctuations,
however they cannot be well described neither within the LLL nor the
XY scaling schemes.

Our analysis shows that despite many similarities with the cuprates
the multi-band nature of the iron pnictides makes them even more complex.

\section{Experimental}

SmFeAs(O, F) single crystals were grown under high pressure in a NaCl/KCl
flux, typically they grew in the form of $5-10\:\mu m$ thick platelets
with $100-200~\mu m$ diameter \citep{Zhigadlo2008}. In the course
of this study we have used two crystals of approximately $40\times50\times5\mu m^{3}$
and $50\times100\times10\mu m^{3}$. The size of the crystals was
estimated from electron microscope images. In order to perform specific
heat measurements on such small samples we have employed membrane
nano calorimeters \citep{Herwaardena} combined with the 345 method
allowing us to measure specific heat of samples as small as 30um in
diameter (corresponding to $\sim50~ng$) \citep{Galeski2016}. After
the specific heat measurements the sample was transferred onto a silicon
wafer and subsequently trimmed into a Hall bar and electrically connected
using the focused ion beam, as shown in the inset of Fig.\ref{Fig. resist}
\citep{Moll2010,Moll2012}. Measurements were performed in a Quantum
Design PPMS cryostat equipped in a 14~Tesla magnet. 

\begin{figure}
\includegraphics[scale=0.46]{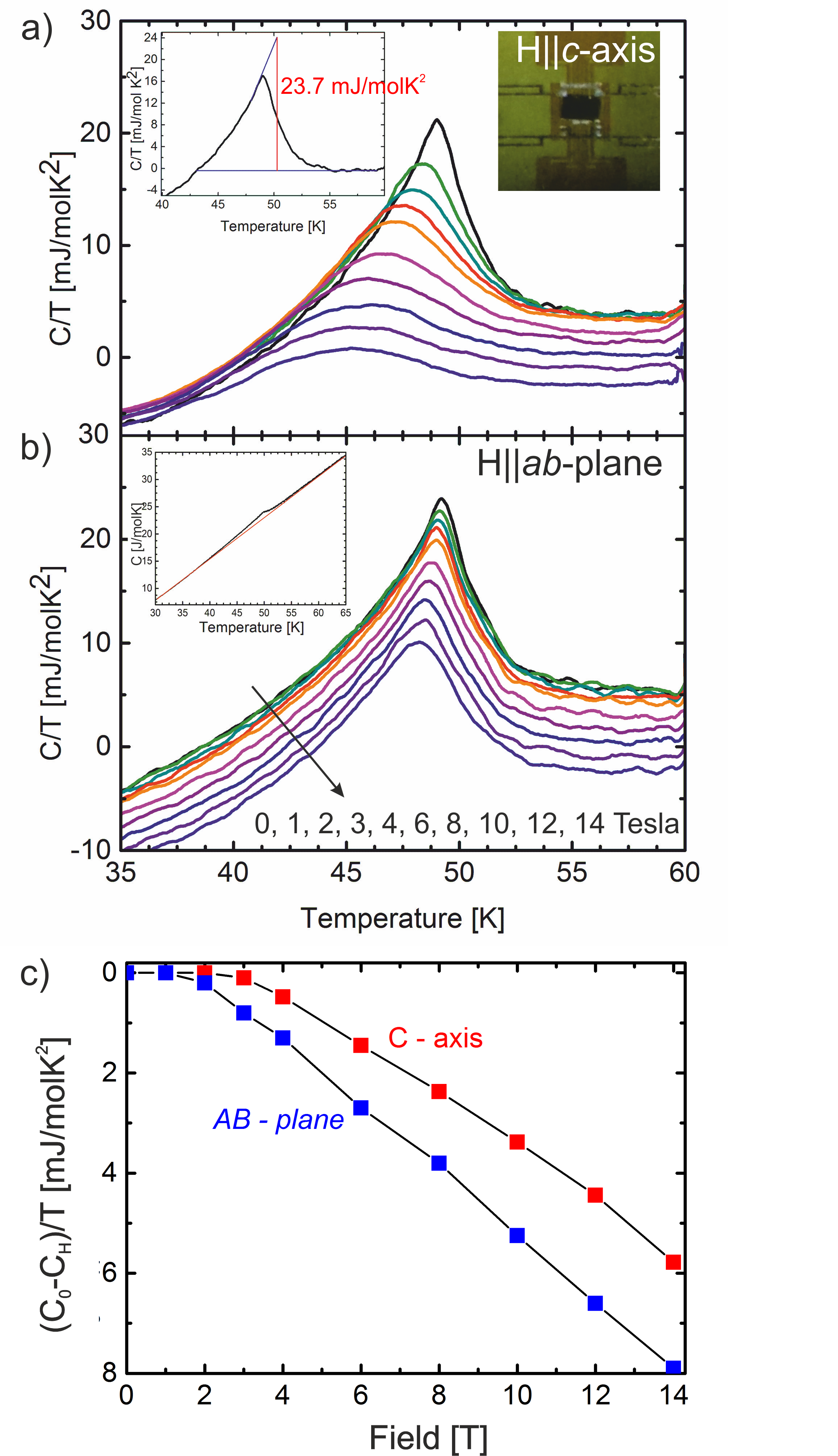}

\caption{\label{fig:Variation-of-the}Variation of the specific heat anomaly
of single crystal SmFeAs(O,F) with the magnetic fields applied along
the \textit{\textcolor{black}{c}}-axis (upper panel) and parallel
to the \textit{\textcolor{black}{ab}}-plane (middle panel). The inset
left of the upper panel exemplifies the procedure used for extracting\textit{\textcolor{black}{{}
T\protect\textsubscript{c} }}at various fields. The inset of the
middle panel presents the total measured specific heat. The bottom
panel shows the field dependence of the difference $C(0T)-C(H).$}
\end{figure}

\section{Specific Heat}

Reliable measurements of specific heat of the '1111' family or in
fact of any crystals that are not of macroscopic size seem to be a
formidable task and there are only a few reports of thermodynamic
bulk measurements performed on nanogram samples \citep{Galeski2016,Welp2008,Welp2011}. 

In order to perform these experiments we have employed membrane based
nano-calorimeters manufactured by Xensor Integration. These chips
although designed for operation in temperatures up to $700\,K$ turned
out to work very reliably at low temperatures down to $1.8\:K$. Unlike
most nano-calorimeters used in condensed matter physics thermometry
in these devices does not relay on resistive thermometers but is based
on a set of 6 compensated silicon thermopiles. Such a design proves
to be especially useful when employing the 345 method as it
allows a direct measurement of the temperature difference between
the sample and the chip frame (For design details please refer to
Xensor technical note \citep{Herwaardena}). 

The superconducting specific heat anomaly amounts to less than 5\%
of the total specific heat (inset of Fig.\ref{fig:Variation-of-the}b).
To extract meaningful thermodynamic information regarding the superconducting
transition we follow the procedure introduced by Welp. et al. \citep{Welp2009,Welp2011}
and subtract a linear background. However for clarity we have subtracted
the same zero field background line from all curves, explicitly showing
the field dependence of the normal state specific heat. (Figure \ref{fig:Variation-of-the},
bottom panel).

Interestingly the normal state specific heat above the transition
is reduced in a magnetic field (Fig.\ref{fig:Variation-of-the}c).
This can be tentatively ascribed to the modification of the crystal-field
split energy levels of the Sm \textit{4f} electrons, as no such suppression
in observed in the Nd- based counterpart \citep{Kanter2015}.

The specific heat near $T_{c}$ is shown in Fig.\ref{fig:Variation-of-the}.
We have fitted a 'mean field jump' (inset of Fig.\ref{fig:Variation-of-the}a)
to the transitions assuming entropy conservation as exemplified in
the inset of Fig.\ref{fig:Variation-of-the}, yielding the following
parameters: $T_{c}=50.5~K$, the upper critical field slopes $H'_{c2}$,
Fig. 1 with the field parallel to the \textit{\textcolor{black}{c-}}axis:
$\sim2.9~T/K$ and $\sim19.5~T/K$ parallel to the \textit{\textcolor{black}{ab-}}plane.
The estimate of the jump hight yielded $\Delta C/T=17.7\:mJ/molK^{2}$
for crystal I and $23.7\:mJ/molK^{2}$ for crystal II.\textcolor{red}{{}
}\textcolor{black}{These values are in fair agreement with data previously
reported by Welp et al. who estimated the anisotropy parameter $\Gamma=\frac{H_{c2}^{ab}}{H_{c2}^{c}}=8$
with the critical field slope along the }\textit{\textcolor{black}{c-}}\textcolor{black}{axis
as $-3.5~T/K$. }

A qualitative investigation of the shape of the specific heat anomalies
reveals a strong superconducting fluctuation contribution with the
high temperature fluctuation tail extending almost 5~K above bulk
$T_{c}$, suggesting some similarities to the cuprate superconductors.
In the case of the cuprates two scenarios were proposed to describe
the behaviour of specific heat in the vicinity of $T_{c}$ in magnetic
fields: the 2D and 3D Lowest Landau Level (LLL) theory and the 3D
XY model \citep{Tesanovic1994,Pierson1995,Jeandupeux,Salamon1993}. 

It was shown that the LLL theory should be a valid approximation for
describing superconducting fluctuations as soon as the magnetic field
becomes strong enough to confine the order parameter to the lowest
Landau level, what translates to a criterion $H>H_{LLL}$ with $H_{LLL}\thickapprox G_{i}H_{c2}(0)$.
The LLL theory predicts that specific heat in the vicinity of $H_{c2}$
should be well described, depending on dimensionality of the fluctuations
by\citep{Tesanovic1994}:

\begin{minipage}[t]{1\columnwidth}%
\medskip{}

\begin{equation}
\frac{dC}{dT}H^{1/2}=F_{2D}^{C}\left(\frac{T-T_{c}(H)}{(TH)^{1/2}}\right)
\end{equation}

\begin{equation}
\frac{dC}{dT}H^{2/3}=F_{3D}^{C}\left(\frac{T-T_{c}(H)}{(TH)^{2/3}}\right)
\end{equation}

\medskip{}
\end{minipage}

where $F_{2D}^{C}(x)$and $F_{3D}^{C}(x)$ are scaling functions.

On the other end the XY model can be considered a justified description
for fields too weak to effectively break the XY symmetry. In this
case the specific heat anomaly is expected to follow the scaling relation
\citep{Jeandupeux,Salamon1993}: 

\medskip{}

\begin{equation}
\left[C(H,T)-C(T,0)\right]H^{\frac{\alpha}{2\nu}}=G\left(\left(\frac{T}{T_{c}}-1\right)H^{\frac{-1}{2\nu}}\right)
\end{equation}

\medskip{}

where $G(x$) is the scaling function and the parameters $\alpha=-0.007$
and $\nu=-0.669$ are the critical exponents characteristic for the
3D XY model.

Interestingly, although both the past and present analysis produce
a similar of value H\textsubscript{c2}, specific heat data extended
to 14~Tesla reveals that the high field data turns out to be not
well described by the 3D Lowest Landau Level (3D-LLL) scaling. 

\begin{figure}[h]
\includegraphics[scale=0.3]{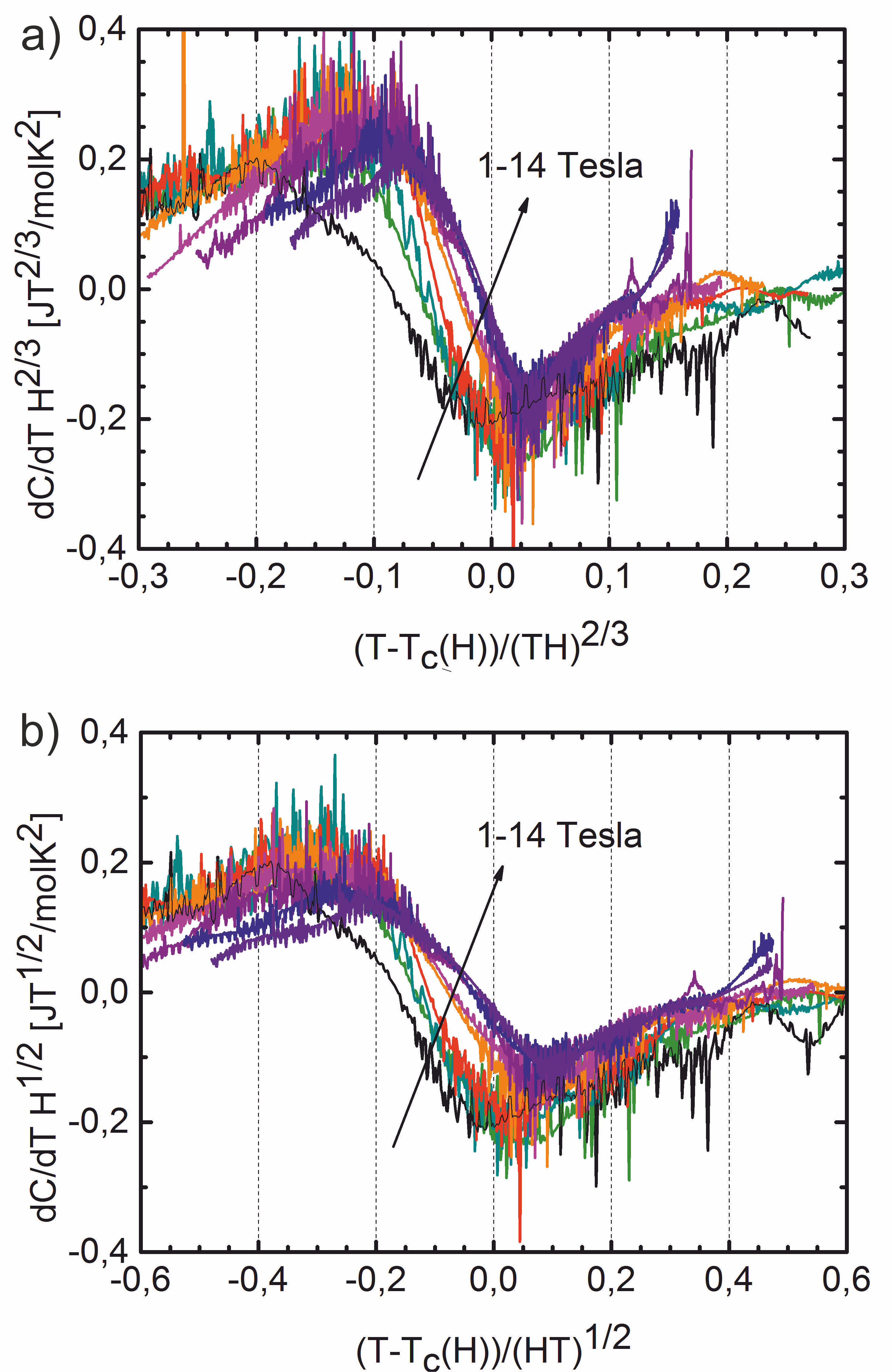}

\includegraphics[scale=0.3]{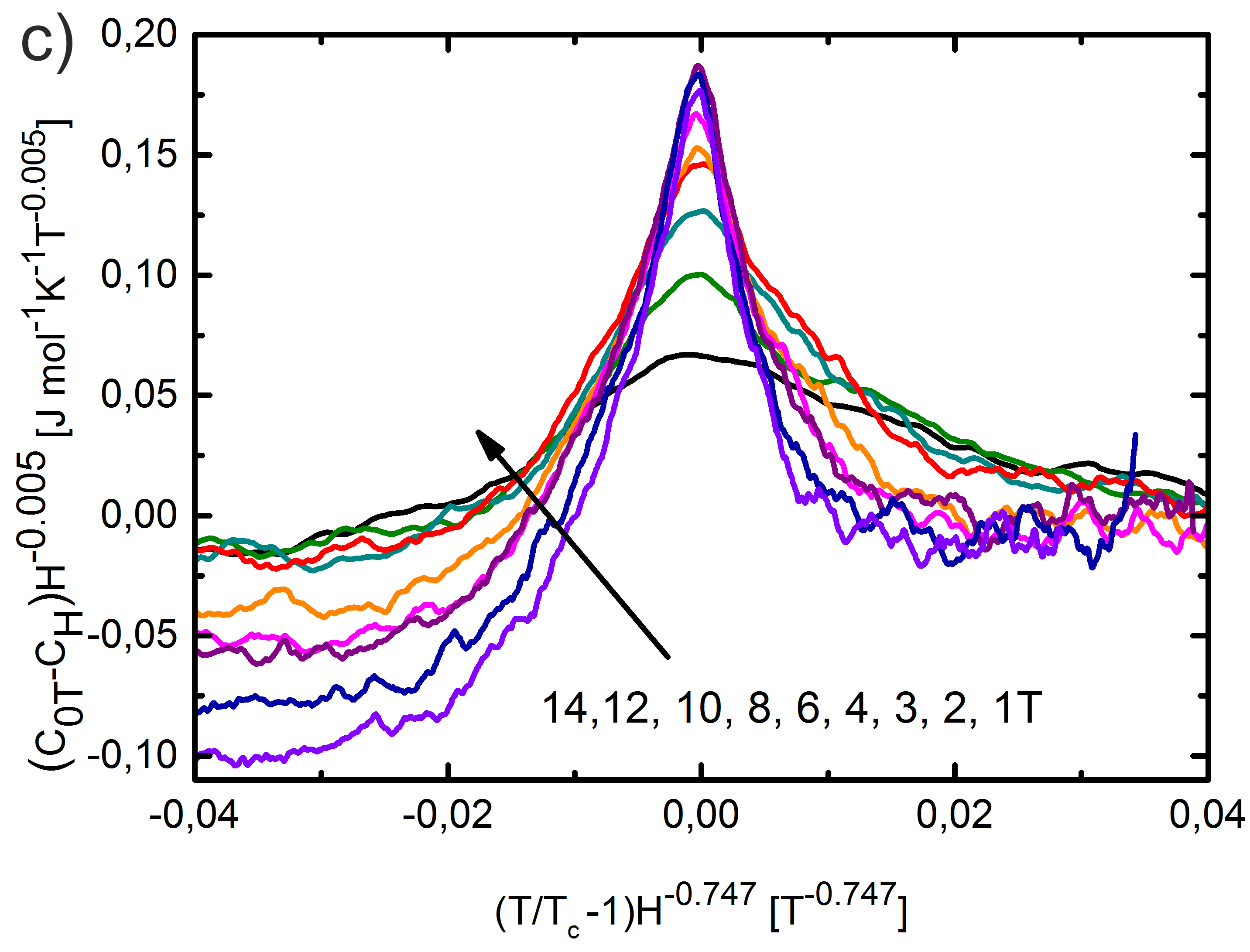}

\medskip{}

\caption{\label{fig:Comparison-of-2D}Comparison of 2D (middle panel) and 3D
(top panel) LLL-scaling schemes with the critical XY scaling (bottom
panel). Best curve collapse for the LLL scaling was obtained for $T_{c}=50.2~K$
and $H_{c2}=-3.1~T/K$ and $T_{c}=49.7~K$ for XY scaling.}
\end{figure}

A close look at Fig.\ref{fig:Comparison-of-2D}a unveils that although
the low field data might suggest an onset of convergence towards 3D-LLL
scaling \citep{Welp2011}, the additional higher field measurements
indicate the 'non-convergence' continues. An attempt at describing
the data using the 2D-LLL scaling (Fig.\ref{fig:Comparison-of-fluctuation}b)
is equally unsuccessful, although it seems to collapse the data slightly
better. A representation of the data scaled within the 3D XY scaling
\citep{Lawrie1994,Jeandupeux} framework (Figure \ref{fig:Comparison-of-2D}c)
is equally unsatisfying. Suggesting that in fact none of the simple
scaling scheme captures all the details of the specific heat anomaly.
This is especially clear when comparing available datasets with near-perfect
data collapse seen in YBCO \citep{Welp1991,Jeandupeux,Overend1996}
or BSCCO \citep{Kobayashi1994}. Additionally comparing the specific
heat anomaly of NdFeAs(O, F) \citep{Welp2008} and SmFeAs(O, F) with
their hydrogenated counterpart \citep{Kanter2015} reveals qualitative
differences in the shape of the specific heat anomalies. This alone
suggest that scaling approaches that proved useful in describing classic
and cuprate superconductors are not sufficient to capture the details
of the physics of the superconducting transition even within one family
('1111') of the pnictides.

\section{Magnetoresistance}

\subsection*{Activated flux flow}

\begin{figure}
\includegraphics[scale=0.3]{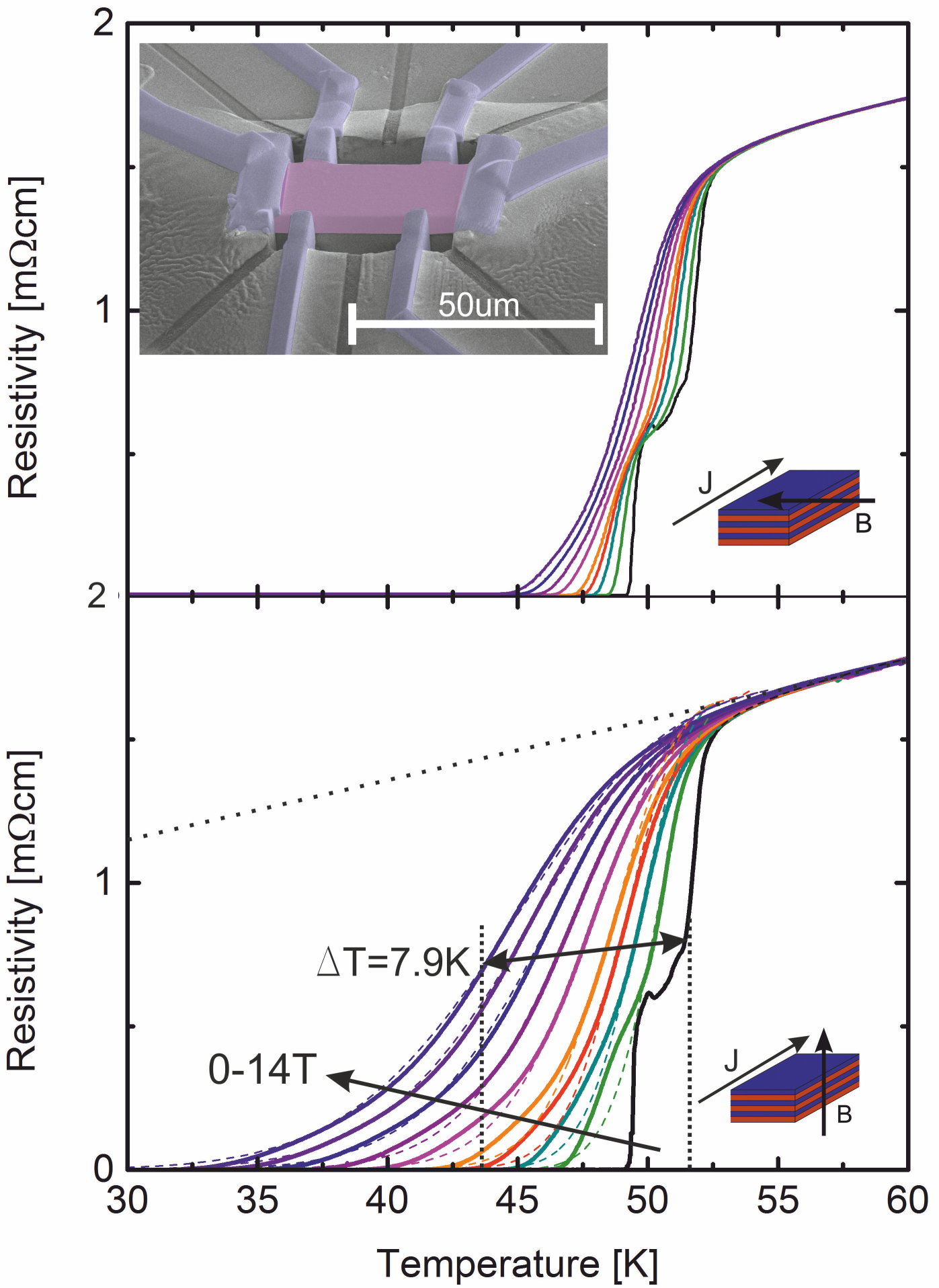}

\caption{\label{Fig. resist}The main panels demonstrate the temperature dependence
of resistivity in magnetic fields along and perpendicular to the \textit{\textcolor{black}{c-}}axis.
The dashed lines in the lower panel are best fits to the activated
flux flow model, see text. The inset shows an electron micrograph
of the sample after preparation for transport measurements. }
\end{figure}

In order to perform electric transport measurements the crystals were
removed from the calorimetric cell, glued to a silicon substrate and
subsequently shaped into a form of a Hall bar and contacted using
the FIB technique \citep{Moll2010}. The measurements were performed
using a $1117.77Hz$ excitation with peak current density of $20~A/cm^{2}$
- well in the Ohmic regime \citep{Lee2010a}, in the same magnetic
fields as the specific heat measurements. The resulting temperature
and magnetic field dependence of the resistive transition is depicted
in Fig~\ref{Fig. resist}. The most prominent feature is the previously
reported broadening of the resistive transition with field applied
along the \textit{\textcolor{black}{c-}}axis. 

It is worthwhile to investigate to what extent these\textit{\textcolor{black}{{}
$\rho(T,H)$}} data can be used to extract the upper critical fields.
So far there were two scenarios proposed in order to describe the
shape of this transition: the first published study, on polycrystalline
samples suggested that the transition width follows the 2D-LLL scaling
relation for fields above 8~Tesla \citep{Pallecchi} . On the other
hand later measurements on single crystal SmFeAs(O, F) demonstrated
the prevailing influence of vortex dynamic and activated flux flow
as determining the broadening of the resistive transition \citep{Lee2010a}. 

Indeed the basic model of activated flux motion as introduced by Tinkham
\citep{Tinkham1988} parametrizes our data surprisingly well in high
fields. For fields below 3\ Tesla (\textit{\textcolor{black}{H║c}})
there is some discrepancy at low temperature , originating from sample
inhomogeneity. This feature becomes unimportant at high fields due
to the intrinsic broadening of the transition and suppression of possible
filamentary superconductivity.

\medskip{}

\begin{equation}
\frac{R}{R_{n}}=\{I_{0}[A(1-t)^{3/2}/2B]\}^{-2}
\end{equation}

\medskip{}

Tinkhams model (eq. 1) describes the phenomenology of the resistive
transition with only two material dependant parameters: T\textsubscript{C}
and $A=CJ_{c0}/T_{c}$ where $C$ can be approximated as $C\approx\beta\,8.07\cdot10^{-3}\,T\,K\,cm^{2}/A$,
with $\beta$ being of the order of unity \citep{Tinkham1988}. 

In the case of SmFeAs(O, F) we have obtained the best fits for \textit{\textcolor{black}{T\textsubscript{c}}}=54K
and the value of $A(H)$ steadily increasing from 120~T and saturating
at 380~T for fields above 4~T. The outcome of the fitting procedure
is represented by the dashed lines in the bottom panel of Figure~\ref{Fig. resist}.
Our analysis displays three remarkable facts: 

(1) The constant $A=\frac{U_{0}}{2T}$ is proportional to the
average vortex activation energy, thus its threefold increase could
be thought of as a manifestation of field dependence of the pinning
potential as suggested by Lee et al \citep{Lee2010a} who found a
transition between two regimes of \textit{\textcolor{black}{$U_{0}(H)$}}
to occur at 3~T. The saturated high field value of $A(H)$, yields
$J_{c0}=2.5\cdot10^{6}A/cm^{2}$ very close to the value found for
YBCO \citep{Tinkham1988}.

(2) The value of $T_{c}$ for which the theory reproduces the data
best is $\sim54\,K$ and remains almost the same for all magnetic
fields. This remarkable fact was already noticed by Tinkham, originally
attributed to the small depression of $T_{c}$ in magnetic field \citep{Tinkham1988}.
Within our framework the value of $54~K$ is significantly higher
than the thermodynamic bulk transition temperature extracted from
specific heat of the same crystal, the investigation of the specific
heat data suggests a clear physical interpretation for the value of
$T_{c}$ used in the activated flux flow model: it is the temperature
defining the onset of superconducting fluctuations. Indeed in SmFeAs(O,
F) this temperature is about $54-55~K$ and seems to be not influenced
by magnetic fields up to 14 T. This opens a question to what degree
the thermodynamic values of \textit{\textcolor{black}{H\textsubscript{c2}}}
and $T_{c}$ are manifesting themselves in the phenomenology of the
resistive transition.

\subsection*{Superconducting fluctuation conductivity}

\begin{figure}[h]
\includegraphics[scale=0.3]{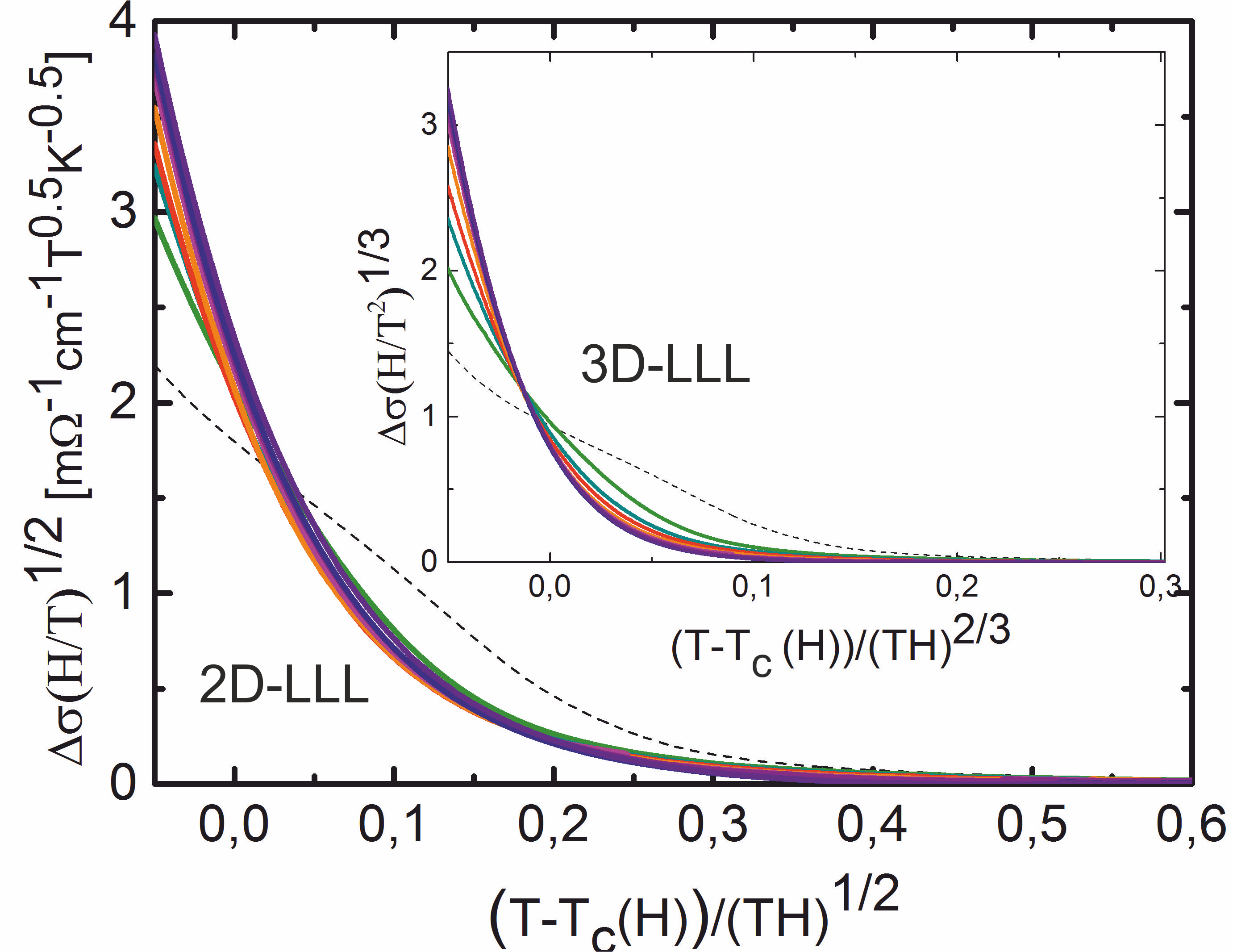}

\caption{\label{fig:Comparison-of-fluctuation}Comparison of fluctuation conductivity
data collapse in 2D (main panel) and 3D (inset) LLL scaling schemes
for fields up to 14T. The dashed line represents the 1T curve, for
details see text. Best data collapse was obtained obtained for T\protect\textsubscript{c}=50.2~K
and H\protect\textsubscript{c2}=-3.1~T/K }

\end{figure}

In the presence of flux motion the activated flux framework describes
most of the shape of the resistive transition remarkably well. However
the first 5-10\% of the drop in resistance are usually dominated by
the presence of superconducting fluctuations above the bulk transition
temperature. To the best of our knowledge the superconducting fluctuation
conductivity has been addressed twice for the '1111' family of superconductors
\citep{Liu2010,Pallecchi}. Pallecchi et al. recognized the shape
of the resistive drop as following a 2D-LLL \citep{Ullah1990,Ullah1991}
scaling relation. However later it was argued that such a result might
have been the effect of using a polycrystalline sample\citep{Welp2011}. 

Similarly as in case of specific heat the LLL theory predicts that
for sufficiently high fields the superconducting fluctuation contribution
to conductivity is expected to follow specific dimension dependant
scaling relations \citep{Ullah1991,Ullah1990}:

\begin{minipage}[t]{1\columnwidth}%
\begin{equation}
\triangle\sigma(H)=\left(\frac{T}{H}\right)^{\frac{1}{2}}F_{2D}^{\sigma}\left(\frac{T-T_{c}(H)}{(TH)^{1/2}}\right)
\end{equation}

\begin{equation}
\triangle\sigma(H)=\left(\frac{T}{H}\right)^{\frac{1}{3}}F_{3D}^{\sigma}\left(\frac{T-T_{c}(H)}{(TH)^{2/3}}\right)
\end{equation}
\end{minipage}

We have extracted the superconducting fluctuation contribution to
conductivity by inverting the resistivity tensor and then subtracting
the extrapolated normal state background. To investigate the dimensionality
of these fluctuations we have plotted the conductivity data in scaled
coordinates. As can be seen in Figure.\ref{fig:Comparison-of-fluctuation}
the two 2D-LLL scaling collapses our data set far better then the
3D scheme: in the 2D case above $4\,T$ (red curve) the collapse is
nearly ideal whereas in the 3D case there is a considerable fanning
out of the curves both below and above $T_{c}$ . What is worth noting
is that the best curve collapse was achieved with $T_{c}=50.2~K$
and $H_{c2}=-3.1~T/K,$ very close the values obtained from the entropy
conservation construction done on the raw specific heat data and from
specific heat scaling. This comparison established the analysis of
the superconducting fluctuation contribution to conductivity as a
rather reliable method to establish $H_{c2}$ and$T_{c}$

\section{Sample quality}

\textcolor{black}{One of the primary concerns when discussing scaling
of superconducting fluctuations is the availability of high quality
crystals. In particular it is essential for this kind of studies to
use single phase crystals, as inhomogeneous substitution throughout
the sample could lead to the appearance of several superconducting
transitions invalidating the scaling analysis. Proper care of this
issue is especially important in materials known to be notoriously
problematic to synthesize, such as the 1111 pnictides.}

\textcolor{black}{Indeed in the case at hand the 'two step' appearance
of the resistive transition (Figure \ref{Fig. resist}) could suggest
a presence of a substantial inhomogeneity in the sample. It is instructive
to compare this two step behaviour with the appearance of the specific
heat anomaly. The specific heat data does not show any signatures
of peak doubling which would be necessarily present for a two phase
sample with two well defined transition temperatures. Additionally
the inspection of magnetoresistance data shows that the first step
in resistivity is suppressed by relatively weak fields suggesting
that the two step appearance of the resistive transition can be attributed
to filamentary/surface superconductivity and can be neglected at high
fields where the LLL- fluctuations scaling should be applicable. }

\section{Conclusions and discussion}

In this study we have investigated the validity of both XY and LLL
scaling schemes in application to a SmFeAs(O, F), a representative
of the pnictide superconductor family with the highest $T_{c}$ .
The analysis revealed that despite structural similarity to YBCO,
SmFeAs(O, F) displays a range of behaviours that cannot be fully accounted
for by theoretical approaches developed for the cuprates. This is
particularly striking when considering specific heat, in YBCO both
LLL and XY scaling are reasonably well describing experimental data
and only considerable experimental effort settled the boundaries of
applicability of both theories. For SmFeAs(O, F) however none of the
approaches provides convincing parametrization of experimental data.
This is especially surprising in the case of LLL scaling. Considering
the relatively low \textit{\textcolor{black}{T\textsubscript{c}}}
and its high Ginzburg number one would expect the LLL theory to be
an adequate description of the condensate in the vicinity of H\textit{\textcolor{black}{\textsubscript{c2}}}
for relatively low fields \citep{Tesanovic1994,Pierson1995,Welp2011}.

\begin{figure}
\includegraphics[scale=0.3]{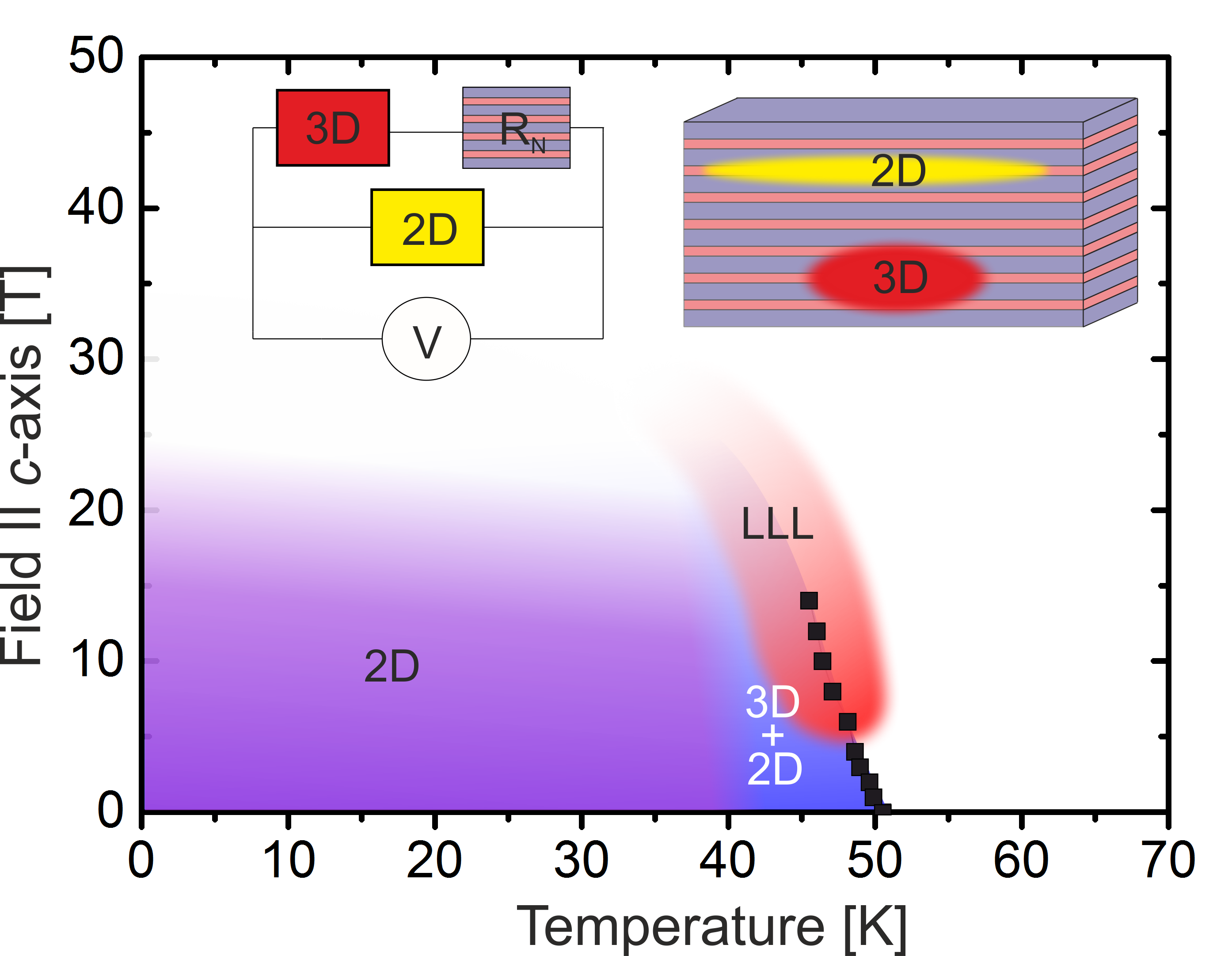}

\caption{\label{fig:phase diagram}A schematic 'Phase diagram' of SmFeAs(O,
F). The insets shows how 2D fluctuations extended in the \textit{ab}-plane
could 'screen' existing 3D fluctuations and lead to the observation
of 2D-LLL scaling.}

\end{figure}

In this context the presence of 2D conductivity fluctuations is even
more puzzling as recent studies of vortex transitions in SmFeAs(O,
F) \citep{Moll2012} conclusively showed that $\xi_{c}\left(T\right)$
remains larger than the inter-plane distance down to $T*\approx0.8T_{c}$
(Fig.\ref{fig:phase diagram}) questioning the 2D nature of superconducting
fluctuations. 

This apparent paradox might be explained by invoking the inherently
multi-band nature of the pnictide superconductors \citep{Kawasaki2008,Szabo,Stanev,Hunte,Gonnelli,Daghero2009}.
In the case of weak inter-band interaction the main contribution to
fluctuation conductivity would come from the band hosting the order
parameter component with the largest $\xi_{ab}$, effectively 'shortening'
all other conductivity fluctuations, inset Fig.\ref{fig:phase diagram}.
If these were of 2D character one could indeed expect the 2D-LLL to
parametrize conductivity well for sufficiently high magnetic fields,
Fig.\ref{fig:phase diagram}. On the other hand specific heat measures
the total change of entropy, and thus would pick up contributions
from both 2D and 3D fluctuations leading to the breakdown of simple
LLL-scaling.

In the case of SmFeAs(O, F) it was shown that two gaps open at the
same temperature with $\Delta_{1}(0)=18$ meV and $\Delta_{2}(0)=6.2$
meV \citep{Daghero2009}. Taking into account the very 2D character
of its Fermi surface, the standard BCS expression for the coherence
length $\xi_{0}=\frac{\hbar v_{F}}{\pi\varDelta}$suggests that
at least one component of the order parameter could indeed be highly
two dimensional.

In this picture at temperatures above $T_{c}$ the properties of SmFeAs(O,
F) are determined by a combination of 2D and 3D fluctuations. On cooling
below bulk $T_{c}$ the 3D component begins to dominate most of the
phenomenology until $\xi_{c}\left(T\right)$ becomes shorter than
the distance between adjacent FeAs layers at which point the dominating
component of the order parameter becomes 2D. 
\begin{acknowledgments}
We would like to thank A. Zheludev, K. Povarov and V. B. Geshkenbein
for many enlightening discussions that helped improve this manuscript. 
\end{acknowledgments}

\bibliography{6Casdas}

\end{document}